\documentclass[11pt]{article}
\usepackage{a4}
\usepackage{amssymb}
\usepackage{array}

\usepackage{graphicx}
\usepackage{hyperref}

\newcommand{\be}{\begin{equation}}
\newcommand{\ee}{\end{equation}}
\newcommand{\bea}{\begin{eqnarray}}
\newcommand{\eea}{\end{eqnarray}}

\newcommand{\eff}{\mathrm{eff}}

\let\nonu=\nonumber

\begin{document}

\title{Renormalization group functions for the Wess-Zumino model:
up to 200 loops through Hopf algebras.}
\author{Marc P. Bellon$^{a,b}$\thanks{On leave from,
Laboratoire de Physique Th\'eorique et Hautes Energies, Boite 126,
4 Place Jussieu,  75252 Paris Cedex 05. Unit\'e Mixte de Recherche
UMR 7589 Universit\'e Pierre et Marie Curie-Paris6; CNRS;
Universit\'e Denis Diderot-Paris7.} and Fidel
A.\ Schaposnik$^{b,a}$\thanks{Associate to CICBA.}\\
 {\normalsize \it $^a$CEFIMAS, Av.\,Santa Fe 1145,
 1069 Capital Federal,
Argentina}\\
{\normalsize\it $^b$Departamento de F\'\i sica, Universidad
Nacional de La Plata}\\ {\normalsize\it C.C. 67, 1900 La Plata,
Argentina}}
\date{\hfill}
\maketitle
\begin{abstract}
We obtain  the contributions to the renormalization group
functions of all the diagrams containing the unique one-loop
primitive divergence of a simple supersymmetric Wess--Zumino
model, up to more than 200 loops. The asymptotic behavior of the
coefficients in the expansion of the anomalous dimension is
analysed.
\end{abstract}

\section{Introduction}

Perturbative Quantum Field Theory is known for its tremendous
successes, with its ability to obtain highly precise values in
Quantum Electrodynamics, or to test the Standard Model with
radiative corrections to weak interactions. However, actual calculations
become rapidly cumbersome and display a conjunction of analytical and
combinatorial difficulties. The fast growth of the
number of relevant terms gets compounded by the need to subtract a growing
number of subdivergences.
%representing $2^n$ different terms for a diagram with $n$.

In the quest of organizing principles for taming the combinatorial
problem, a major progress has been the recognition of a Hopf
algebra structure in the renormalization of Quantum Field Theories
\cite{kreimer1,CK1,CK2,CK3}.
The cohomology of the introduced Hopf algebras has been related to
Schwinger--Dyson equations (see~\cite{Krev} and the references therein).
An early application to the summation of a
category of diagrams in a simple Yukawa field theory or a $\phi^3$
theory in 6 dimensions has been obtained by Broadhurst and Kreimer
in~\cite{BrKr02}. It was however later shown that in this simple case,
the Schwinger--Dyson equation, which is linear, could be solved exactly, without any
reference to the Hopf algebra of diagrams, since it gives rise to a
non-linear differential equation for the anomalous dimension~\cite{BrKr01}.

In a preceding work~\cite{BeLoSch}, we adapted this work to
a supersymmetric model with a simple Wess--Zumino type interaction.
As it is well-known, in view of its ability to limit renormalization
phenomena, supersymmetry has allowed to resolve a number of
theoretical problems. Although its relevance as a property of the
physical world is up to now as elusive as ever, it is still a
leading candidate for physics beyond the standard model. Of course,
if supersymmetry plays a role in particle physics, it has to be
broken through some (preferably spontaneous or dynamical) mechanism.
In this respect, theories with metastable supersymmetry breaking
have been recently very actively investigated, particularly some
generalized Wess--Zumino models~\cite{ISS1,ISS2}.

% Starting precisely from a  supersymmetric Wess--Zumino model with
% a unique primitive divergence, we extended in \cite{BeLoSch} the
% renormalization analysis of Broadhurst and Kreimer~\cite{BrKr01}
% taking profit of the simplifications
% induced by supersymmetry.
The method employed in~\cite{BrKr01}, if really powerful, does not
generalize to more complex situations. A strategy has been proposed
in~\cite{KrYe1} to deal with non-linear Schwinger--Dyson, in which
the formulation of the renormalization group in terms of the Hopf
algebra of diagrams becomes essential. We here apply this proposal
in a supersymmetric setting. With respect to the cases envisaged
in~\cite{KrYe1}, the supersymmetric case is simpler because we need
only consider a unique anomalous dimension, which is common to all
the member of the supermultiplet. We further simplify the procedure
by a direct computation of the renormalized propagator, so that we
can obtain explicit results up to high orders.

We will first review the bases of this procedure, which are the
following two results: (i) the renormalization group is a
one-parameter subgroup of the group of evaluation functions on the
Hopf algebra of graphs~\cite{CK3} and (ii) some combinations of
graphs form sub-Hopf algebras.  We will show that these results
allow to recover the full propagators from the renormalization group
functions.

We then apply this calculation scheme to the Wess--Zumino supersymmetric
model, building on our preceding work~\cite{BeLoSch}. As we shall see, the
efficiency of the method allows to reach high degrees of perturbation
theory, being essentially limited by the size of the expression in terms of
values of the Riemann $\zeta$ function.  Switching to numerical
approximations allows to go further and reach the asymptotic behavior of
the coefficients of the development of the anomalous dimension $\gamma$ in
powers of the ``fine coupling constant'' $a$. This behavior allows for the
definition of a convergent Borel transform of this serie. A singularity is
however present on the positive axis but we discuss the options for
nevertheless obtaining a sensible resummation.  This work suggests further
developments that we explore as a conclusion.

\section{Exponentiation of the renormalization group}

Our purpose is to reach high order of perturbation theory in the
evaluation of renormalization group function. Along the lines suggested
in~\cite{KrYe1}, we use Schwinger--Dyson equations to produce a family of
diagrams for which  analytical evaluation is very simple. However, a na{\"\i}ve
approach to this evaluation involves a triple serie in the coupling
constant $g$, the logarithm of the momentum $L=\log(q^2/\mu^2)$ and the
dimensional regularization parameter $\epsilon$.  It was
shown  in~\cite{BrKr01} how to reduce the problem
 to a single non--linear differential equation for the
renormalization group $\gamma$-function, but the proposal relied on the
linearity of the Schwinger--Dyson equation.

Now, if we work only with renormalized quantities, we can avoid the
$\epsilon$-parameter. The momentum dependence can then be recovered
from the renormalization group. Consider  the ratio of the full
propagator with the free propagator which,  in a Lorentz invariant
scheme and for a massless theory,  is a single number. It can be
fixed to 1 at the scale $\mu^2$ in a ``physical'' renormalization
scheme and the value for any momentum $p$ can be inferred  from the
action of the renormalization group with parameter
$t=\log(p^2/\mu^2)$.  Other renormalization schemes like the minimal
substraction one or its variants would introduce finite
contributions at the reference momentum. Similarly, vertex
functions, which generically depend on multiple scales, would
involve finite contributions at the reference scale. However, here
we are interested only in the propagator of a massless theory where
these complications are not present and the full propagator can be recovered
from the renormalization group.

In~\cite{CK3}, it was shown that the renormalization group is a
one parameter subgroup $\{F_t\}$ of the group of characters of the
Hopf algebra of graphs, which means:
\begin{eqnarray}
F_t(XY) &=&  F_t(X) F_t(Y)\nonumber
\\
 \label{group}
    F_{s+t}( X ) &=& (F_s\otimes F_{t})(\Delta X)
\end{eqnarray}
Multiple derivations of this identity allows to recursively obtain the
derivatives of $F_t$ at the origin.
Indeed:
\begin{equation}\label{der+1}
\partial_t^{\,k+1} F_t\bigr|_{t=0} = \partial_s^{\vphantom{k}} \partial_{t}^{\,k}
F^{\vphantom{k}}_{s+t}\bigr|_{s=0,t=0} = \partial_s F_s\bigr|_{s=0}\,
\partial_{t}^{\,k} F^{\vphantom{k}}_{t}\bigr|_{t=0}
\end{equation}
Applied on a group-like element of the Hopf algebra, that is, on
an element $X$ such that $\Delta X = X \otimes X$, this identity
shows that the successive derivatives are powers of the first one,
so that $F_t(X)= e^{\alpha t}$. We will deal however with more
general elements of the Hopf algebra of diagrams, so that the
results will be more complicated.

Finally since $F_t$ is a character, its evaluation on a product is the product
of the evaluations, so that $\beta= \partial_t F_t \vert_{t=0}$, the generator
of the renormalization group, is a derivation. This will be important for
the applications of eq.~(\ref{der+1}).

\section{Hopf algebra of Green functions}

In order to make the best use of the Hopf algebra technique, it is
convenient
 to work not at the level of the individual diagrams, but on some sums of
diagrams. If the sums of interest can be shown to form sub Hopf algebras,
this allows to overcome in part the combinatorial complexities of
calculations.

Our starting point is that the effective coupling is a map from the
Hopf algebra of the graphs to the Hopf algebra of formal
diffeomorphism tangent to the identity in 0. For couplings
associated to a three-point vertex, it will be convenient to express
this fact not in terms of the coupling $g$, but of a ``fine
structure constant'' $a=C{g^2}$, in order to avoid square roots. The
proportionality constant $C$ can be choosen as in QED to simplify
further results. If the vertex couples particle species $j$, $k$ and
$l$, the effective constant is defined as:
\begin{equation}
a_\eff = \frac { C \Gamma_v^2 } { \Gamma_{2,j} \Gamma_{2,k}
\Gamma_{2,l} }
\end{equation}
Here $\Gamma_{2,j}$ is the two-point function for the $j$ specie
while $\Gamma_v$ is the effective interaction vertex.

 Writing
\begin{equation}
a_\eff = \sum_n a_{\eff,n} a^n
\end{equation}
the action of the coproduct on $a_\eff$ is given by~\cite{CK3,vSl}
\begin{equation} \label{deltaA}
 \Delta a_\eff = \sum_n a_\eff^n \otimes
 a_{\eff,n}
\end{equation}
This corresponds to the fact that $a_\eff$ is a map to the Hopf algebra of
(coordinates for) formal diffeomorphisms.

In fact, two refinements are easy to obtain. First, the same formula holds
whenever we do not take for the $\Gamma$'s the full perturbative expansion,
but  only the terms which contain a certain set of primitive divergences.
This holds in particular if we take for the $\Gamma$'s the approximation
resulting from a given truncation of the Schwinger--Dyson equations.

An examination of the proof of the previous identities
in~\cite{vSl} shows that  stronger results hold  for  the
two- or three-point functions which are involved,
\begin{eqnarray}
    \Gamma_{2,m} &=& \sum_n \Gamma_{m,n} a^n, \\
    \Gamma_v &=& g \sum_{n=0} \Gamma_{v,n} a^n, \\
    \Delta\Gamma_{2,m} &=& \sum_n \Gamma_{2,m}a_\eff^n \otimes
    \Gamma_{m,n} \label{deltaG2}\\
    \Delta\Gamma_v &=& \sum_n \Gamma_v a_\eff^n \otimes
    \Gamma_{v,n}\label{deltaGv}
\end{eqnarray}
In the cases in which the coupling constant is not renormalized,
$a_\eff=a$, these equations indicate that the $\Gamma$'s are group
like elements of the Hopf algebra.  Now, when deriving
relation~(\ref{deltaA}) from
relations~(\ref{deltaG2},\ref{deltaGv}), one uses the fact that
formulas of this kind are multiplicative. Indeed, suppose that
$A=\sum A_n a^n$ and $B=\sum B_n a^n$ are two series satisfying a
property like~(\ref{deltaG2}):
\begin{equation} \label{gener}
 \Delta A = \sum_n A\, a_\eff^n \otimes A_n \, .
\end{equation}
Then the product $AB$ satisfies the same property. We can further
remark that the coproducts of the propagators satisfy similar
equations. This easily follows from the fact that the identity
element of the Hopf  algebra trivially satisfies~(\ref{gener}). This
allows to prove~(\ref{deltaA}) from
eqs.~(\ref{deltaG2},\ref{deltaGv}).

\section{Practical formulas}

Joining the results of the two preceding sections allows to
construct the full renormalized propagator at any order in
perturbation theory from the first derivative with respect to $L$,
that is, the renormalization group function $\gamma$. In fact, we
can get at will the propagator or its inverse, the 1PI two-point
function.

Let us consider the evaluation of the renormalization group on the
propagator $\Gamma_{2,j}^{-1}$, normalized by the free propagator,
\begin{equation}
F_t(\Gamma_{2,j}^{-1})= 1+\sum_{n=1}^\infty \gamma_k t^k \,
\label{eleven}
\end{equation}
where, as before,  $t=\log(p^2/\mu^2)$.  Notice that the coefficient
$\gamma_1$ in expansion (\ref{eleven}) is related to the anomalous
dimension $\gamma$,
\begin{equation}
\gamma_1 = -2 \gamma
\end{equation}
The successive terms of the development of this expression in
powers of $t$  can be computed from eq.~(\ref{der+1}), using the
coproduct in eq.~(\ref{deltaG2}),
\begin{equation}
(k+1) \gamma_{k+1} =\sum_n  (\gamma_1 + n \beta)a^n \gamma_{k,n}  =
(\gamma_1 + \beta \, a \partial_a) \gamma_k \label{nom}
\end{equation}
To obtain  the preceding equation we   made use  of the simple differentiation rule given by~(\ref{der+1}). This allowed to apply the derivation
property of the generator of the group $ \partial_t F_t\bigr|_{t=0}$
to further evaluate its effect on the left factor of $\Delta A$.

\section{Application to the Wess--Zumino model}

The model we will consider is the simplest Wess--Zumino model, with
only one type of superfields (and not two as in our preceding
work~\cite{BeLoSch}). The general non--renormalization properties of
such a model ensures that the vertex is not renormalized and in
fact, in a massless theory, it cannot get any correction. The only
contribution to the renormalization group comes from the correction
to the propagator and the supersymmetry has here the  effect that
all components of the superfield get the same renormalization
factor\footnote{In a supersymmetric gauge theory, the situation
would be subtler, since the gauge symmetry and the supersymmetry are
not easy to combine: in the component formalism, supersymmetry is
not fully explicit and in a superfield formalism, the unconstrained
superfield describing the gauge supermultiplet is of dimension zero,
so any power of this field can appear in the renormalized
Lagrangian. This can be overcome by an appropriate choice  of the
gauge fixing condition and therefore it does not affect physical
properties~\cite{Piguet}.}. We therefore have the simple relation
\begin{equation}
    \beta = 3 \gamma_1.
\end{equation}
so that eq.(\ref{nom}) becomes
\begin{equation}
(k+1) \gamma_{k+1} =  \gamma_1(1 +3 a \partial_a) \gamma_k
\label{nome}
\end{equation}

The first step in our calculation is to compute the effect of the
one--loop primitive divergence. There are no primitive divergences
at two loops and the three--loop one is non-planar and therefore
subleading in a large $N$ study. For this we will insert the full
renormalized propagator at order $n$ in the one loop diagram to
derive $\gamma_1$ at the order $n+1$. Since $\gamma_k$ is at least
of order $k$, we only have to consider a finite expansion in $t$ of
the correction factor of the propagator.

We therefore need to compute a simple one loop integral, but with
propagators having logarithmic corrections. All of them can be
obtained multiplying the propagators $\Gamma_2^{-1}(p^2)$  by a
factor $(p^2/\mu^2)^x$. Indeed, differentiating with respect to $x$
at $x=0$ will give the logarithm factors  in each of the
propagators. This procedure is essentially  the same as the Mellin
transform method used in~\cite{KrYe1}.

As in~\cite{BeLoSch}, we only need to compute the graph for the
correction to the auxiliary field propagator. Indeed, the fact that
we have corrections on the two propagators does not change the
observation already made there: the corrections to the free
propagators of all the members of the supermultiplet are equal. We
are interested in the quantity

\begin{eqnarray}\label{Mellin}
    \Gamma(q^2,x,y) &=& + \frac{g^2}{8\pi^4}\int d^4p
    \frac{1}{(p^2)^{1-x}[(q-p)^2]^{1-y}}\nonu\\
    &=&+\frac{g^2}{8\pi^2\Gamma(1-x)\Gamma(1-y)}
    \int du \,dv\, \frac{u^{-x} v^{-y} }{(u+v)^2} \exp\left( -
    \frac{uv}{u+v}q^2\right)\nonu\\
    &=& + \frac{g^2}{8\pi^2} (q^2)^{x+y} \frac{\Gamma(-x-y)\Gamma(1+x)\Gamma(1+y)}
        {\Gamma(2+x+y)\Gamma(1-x)\Gamma(1-y)}
\end{eqnarray}
The overall plus sign comes from the product of the minus sign in
every correction to the 1PI two-point function and the free
propagator which is $-1$. It will be convenient to introduce the
``fine structure constant'' $a=g^2/8\pi^2$. Now, we can get an
expression similar (but simpler) to the ones in~\cite{KrYe1} using
the expansion of the logarithm of the $\Gamma$ function
\begin{equation}\label{Mellin2}
    \Gamma(q^2,x,y) = - \frac{a (q^2/\mu^2)^{z}}{z(1+z)}
    \exp\left( 2\sum_{l=1}^\infty \frac{\zeta({2l+1})}{ 2l+1}\bigl(
    z^{2l+1} - x^{2l+1} -y^{2l+1}\bigr) \right)
\end{equation}
where $z=x+y$ and $\zeta({2l+1})$ is the Riemann zeta function.
From~(\ref{Mellin2}) we can obtain the two-point functions
calculated with modified propagators by applying differential
operator in $x$ and $y$. Differentiation with respect to $\log(q^2)$
produces a factor $z$ which exactly compensates the term which is
singular for $z=0$. Let us define
\begin{equation}
H(x,y) = -\left.\frac{\partial \Gamma (q^2,x,y)}{\partial \log(q^2)}
\right\vert_{q^2 = \mu^2}
\end{equation}
Then, one has
\begin{equation} \label{Mellin3}
H(x,y) = \frac{a}{1+x+y}\exp\left( 2\sum_{l=1}^\infty
\frac{\zeta({2l+1})}{ 2l+1}\bigl(
   (x+y)^{2l+1} - x^{2l+1} -y^{2l+1}\bigr) \right)
\end{equation}
and
\begin{equation}
 \gamma_1 = (1+\sum \gamma_n \frac{d^n}{dx^n})(1+\sum \gamma_n
 \frac{d^n}{dy^n}) H(x,y) |_{x=y=0} \label{final}
\end{equation}

The evaluation of the $\gamma$ function does not require any further
regularization. After the order $a$ term, which is obtained making
$x=y=0$, the following terms involve corrections to the propagators,
which  are obtained by taking derivatives with respect to $x$ and
$y$ evaluated at the origin, this corresponding to diagrams with the
propagators multiplied by given powers of $\log(p^2)$.

We could
suppose that all the formalism we developed is superfluous and that
we could simply derive further  eq.~(\ref{Mellin2}) with respect to
$\log(q^2)$ to obtain directly the higher terms in the expansion.
Although this is possible, the computation are really much more
cumbersome, even for the leading term of the derivatives of
$\Gamma_2$. Furthermore, one would obtain the development of the
2-point function in the effective action, so that an inversion of
the expansion would be needed to obtain the propagator.

\begin{table}
\setlength{\extrarowheight}{5pt}
\[ \begin{array}{c}
\gamma(a)= a -2 a^2 + 14 a^3 + \Bigl( -160 +16\,\zeta(3) \Bigr)
a^{4} \\ + \Bigl( 2444 -328\,\zeta (3) \Bigr) a^{5} + \Bigl( -45792
+7056\,\zeta(3) +2016\,\zeta(5) \Bigr) a^{6} \\ + \Bigl( 1\,005\,480
-169\,152\, \,\zeta(3) -70\,896\,\,\zeta(5) +8960\,\zeta^2(3) \Bigr)
a^{7} \\+ \Bigl( -25\,169\,760 +4\,509\,408\,\,\zeta (3)
+2\,199\,840\, \,\zeta(5) \\
+564\,480\,\,\zeta(7) -390\,400\,\,\zeta^2(3) \Bigr) a^{8} \\+
\Bigl( 705\,321\,200 -132\,548\,640\,\,\zeta(3 )
-69\,922\,848\,\,\zeta(5 ) \\
-29\,005\,632\,\,\zeta(7) +14\,193\,504\,\,\zeta^2(3)
+6\,397\,056\,\, \zeta(3)\,\zeta(5) \Bigr) a^{9} \\ + \Bigl(
-21\,841\,420\,384 +4\,261\,047\,424\,\,\zeta (3)
+2\,354\,993\,856\,\,\zeta(5) \\
+1\,194\,909\,696\, \,\zeta(7) +{{858\,457\,600}\over{3}}\,\zeta(9)
-512\,441\,536\,\,\zeta^2(3) \\
-383\,788\,416\,\zeta(3)\,\zeta(5 )
+{{49\,556\,480}\over{3}}\,\zeta^3(3 ) \Bigr) a^{10} \\ + \Bigl(
740\,194\,188\,032 -148\,784\,410\,432\,\,\zeta(3)
-84\,779\,661\,888\,\,\zeta(5 ) \\
-47\,818\,582\,272 \,\zeta(7)
-{{58\,999\,853\,440}\over{3}}\,\zeta(9)
+19\,225\,297\,088\,\, \zeta^2(3) \\
+17\,828\,697\,216\, \,\zeta(3)\,\zeta(5)
+1\,829\,076\,480\,\zeta^2(5) \\
+3\,838\,602\,240\,\,\zeta(3)\,\zeta(7)
-{{3\,432\,237\,056}\over{3}}\,\zeta^3(3) \Bigr) a^{11} \\ + \Bigl(
-27\,243\,674\,154\,368 +5\,610\,375\,120\,768\,\,\zeta(3)
+3\,266\,192\,145\,024\,\,\zeta(5 ) \\
+1\,961\,976\,190\,464 \,\,\zeta(7)
+1\,019\,076\,124\,160\,\,\zeta(9 )
+230\,546\,534\,400\,\,\zeta(11) \\
-760\,702\,109\,184\,\,\zeta^2 (3) -788\,057\,929\,728\,\,\zeta
(3)\,\zeta(5)
-141\,297\,435\,648\,\,\zeta^2(5) \\
-297\,887\,016\,960\,\,\zeta(3)\,\zeta(7)
+59\,550\,068\,736\,\,\zeta^3(3)\\
+34\,512\,334\,848\, \,\zeta^2(3)\,\zeta(5) \Bigr) a^{12} + \cdots
\end{array} \]

\caption {\small The first 12 terms of the development of
$\gamma_1$} \label{gamma}
\end{table}

\section{Calculation and results}

Once the Mellin transform~(\ref{Mellin3}) is known, it is
straightforward to obtain the first orders in the development of the
$\gamma$ function from eqs.~(\ref{nome}) and~(\ref{final}). We give
in table~\ref{gamma} the 12 first terms of the expansion of
$\gamma$. The principal limitation for calculating  this ``exact"
form for each term stems from the rapid growth, which behaves as
$\exp(\alpha n^{2/3})$, of the number of contribution one has to
take into account. Already in the terms presented in
table~\ref{gamma}, one can see the rapid growth of their size. At
order 25, there are already 122 possible products of $\zeta$ values
contributing to the result and this number reaches 409174 at order
100. Such a number, as well as the size of the corresponding
rational coefficients, limit to less than 30 the number of terms
which can be computed with a straightforward implementation on
today's average computer, mainly due to the memory footprint. The
rapid growth of the required resources limit to a few units the
number of additional terms which can be computed using more
efficient coding schemes and bigger memories.  The highest orders of
the development cannot therefore be reached in this way, all  the
more  so  that at the end we will need to convert these results to
some concrete numerical approximation.

\begin{table}
\[  \setlength{\extrarowheight}{4pt}
\begin{array}{c}
\gamma(a) =a
-2\;a^2
+14\;a^3
-140.767089549446491434 \;a^4 \\
+2049.72533576365307439  \;a^5
-35219.8401369368689507 \;a^6 \\
+741582.310142069315875 \;a^7
-1.74630317191742523615 \! \times \! 10^{7}\;a^8 \\
+4.72719801334671229530  \! \times \! 10^{8}\;a^9
-1.39759545666280992694 \! \times \! 10^{10}\;a^{10} \\
+4.60146704077682933925 \! \times \! 10^{11}\;a^{11}
-1.63220296094286720854 \! \times \! 10^{13} \;a^{12} \\
+6.32651854893093835423 \! \times \! 10^{14}\;a^{13}
-2.61715263667021333524 \! \times \! 10^{16}\;a^{14} \\
+1.16791189443603376676  \! \times \! 10^{18}\;a^{15}
-5.52247245848724267096 \! \times \! 10^{19}\;a^{16} \\
+ \cdots \\
+8.4053176185682527526 \! \times \! 10^{454}\;a^{195}
-4.9339110330514367678 \! \times \! 10^{457}\;a^{196} \\
+2.9112362346747106444 \! \times \! 10^{460}\;a^{197}
-1.7263592738217495952 \! \times \! 10^{463}\;a^{198} \\
+1.0289894774008300571 \! \times \! 10^{466}\;a^{199}
-6.1636327768018535021 \! \times \! 10^{468}\;a^{200} \\
+3.7107878544109289292 \! \times \! 10^{471} \;a^{201} + \cdots
\end{array}
\]
\caption {\small Approximations of the development of $\gamma_1$}
\label{gamma_n}
\end{table}

It is therefore convenient to work from the start with numerical
values. A much higher number of terms is now easily within reach,
since for a fixed precision, the computation work grows only
quartically with the desired number of terms. In
Table~\ref{gamma_n}, we give the development of $\gamma_1$ up to 16
loops and the 7 last terms we calculated. The full table, with all
results up to 201 loops, is available upon request from the authors.

In the calculation of the expansion of $\gamma_1$, we first have to
determine the Taylor expansion of the  function $H(x,y)$. This is
the part of the calculation which is numerically most demanding,
since these coefficients are the sum of a great number of terms
which nearly cancel, so that care must be taken that the true
precision of the result is lesser than the number of digits used in
the computation. However, something rather remarkable happens: the
Taylor expansion of the rather complicated function $H$
in~(\ref{Mellin3}) is, asymptotically,  the same as the one of $h$,
defined by:
\begin{equation}
h(x,y) = (1+xy)\left(\frac1{1+x}+\frac1{1+y}-1\right) +
{\textstyle\frac12}\frac {xy}{1-x-y} \label{asymp}
\end{equation}
More precisely, the terms of total degree $n$ in the Taylor expansion of
$h$ and the original function $H$ are the same up to a relative error which is
smaller than $2^{-n}$. If it is relatively easy to understand the
properties of the terms in $x y^n$ from the serie with general term
$\zeta(2p+1)-1$, the interplay of different terms which yields such a
simple result for the general terms appears quite miraculous.  Even more,
the error term proportional to $2^{-n}$, $3^{-n}$ and $4^{-n}$ can be
given similar explicit forms and the determination of error terms
proportional to powers of bigger integer seem only to require more terms
with higher precision. These asymptotic formulas have been used to obtain
the higher degrees of the Taylor expansion for our numerical application.

The asymptotic behavior of the coefficients $g_n$ in the expansion
of $\gamma_1$ in powers of $a$ can be determined. With
\begin{equation}
\gamma_1(a) = \sum_{n=1}^\infty g_n a^n,
\end{equation}
we have that
\begin{equation}
    g_n \propto (-3)^n \Gamma(n+2/3).
\end{equation}
 This implies that
a suitably defined Borel transform will have a single pole at $-1/3$
as dominant singularity. However the asymmetry between odd and even
terms of the series implies that there
  is also a weaker singularity at $+1/3$.
It is not a simple pole but the singularity of the dilogarithm ${\rm
Li}_2$.
 This precludes the analytic continuation of
the Borel transform on the positive real axis and its approximation
by Pad\'e methods. A brutal answer to this problem is simply to
integrate the Borel transform only on the interval $[0,\frac13]$.
This gives an analytic function which has the desired asymptotic
expansion at the origin. However this procedure is rather ad hoc and
gives a  far from unique solution.

There is also the possibility that higher order corrections to
the Schwinger--Dyson equation stemming from the addition of multiloop primitive
divergences would weaken the singularity on the positive axis, so that
the singularity on the positive axis would disappear in a complete
calculation.

Taken as an asymptotic formula, eq.(\ref{asymp}) could be useful for
a demonstration of the asympotic properties of the expansion of
$\gamma_1$. The second term in $h$ has positive Taylor coefficients.
Combined with the alternating signs of the main term in the
development of $\gamma_1$, this introduces strong compensation
between the terms with even total powers of derivatives in
eq.~(\ref{final}) and those with odd total powers. As a result, in
the calculation of the term of degree $n$, the terms with one of the
propagator with the  correction of degree $n-1$ and the other one
equal to the free propagator are dominant.

\section{Discussion}

 High orders of the perturbative development of the
renormalization group function have been obtained, resumming all
diagrams with the simplest primitive divergence. The number of terms
calculated in the corresponding expansion allows to reach a clearly
asymptotic regime. The Borel transform of the resulting series
however presents singularities on the positive real axis. The
question remains open whether this signals a fundamental limitation
of the perturbative calculations or   it is just an artifact of the
present approximation. Indeed, in a large $N$ approximation, other
primitive divergences are possible, beginning at the fourth order.

However, one should first test if the additional contributions
remain subdominant. Indeed, in the contribution from the four loop
primitive divergence, the corrections to the propagators can be
split between the eleven propagators, generating a rapidly growing
number of terms and this could more than compensate for the fact
that corrections to the propagator of order $n-3$ are of order
$1/n^3$ with respect to the propagator of order $n$ which gives the
main contribution in the term we have calculated. This would also be
an important step in the way of conjecturing $\zeta$-function style
functional equations generalizing the one obtained in~\cite{KrYe1}
for the simple case where the $\gamma$ function satisfies a
differential equation.

One interesting feature of the approach we presented is that it
provides a clear path for evaluating the correction stemming from
additional terms in the Schwinger--Dyson equation. The evaluation of
the primitive diagrams looks formally as the analytic regularization
proposed long ago~\cite{BoGiGD64,Sp68,Sp71} but the fact that we
need such an evaluation only for primitively divergent diagrams
avoids much of the technicalities stemming from the complex
multidimensional pole structures appearing in general diagrams.

\vspace{1 cm}

\noindent\underline{Acknowledgments}

\noindent We would like to thank the Sociedad Cientifica Argentina
for hospitality during part of this work.
 This work is partially supported by CONICET (PIP6160), ANPCyT (PICT 20204),
 UNLP,  and CICBA  grants. M.B acknowledges CNRS support
 through his ``mise \`a disposition".

\end{document}